\documentclass[aps]{revtex4}

\usepackage{verbatim,graphicx,float}
\usepackage[official]{eurosym}

\begin{document}

\title{Bayesian reasoning in cosmology}

\author{Jakub \surname{Mielczarek}\email{jakub.mielczarek@uj.edu.pl}}
\affiliation{Astronomical Observatory, Jagiellonian University, 30-244
Krak\'ow, Orla 171, Poland}
\affiliation{Copernicus Center for Interdisciplinary Studies,
30-387 Krak\'ow, Gronostajowa 3, Poland}

\author{Marek \surname{Szyd{\l}owski}\email{uoszydlo@cyf-kr.edu.pl}}
\affiliation{Department of Theoretical Physics,
The John Paul II Catholic University of Lublin,
Al. Rac{\l}awickie 14, 20-950 Lublin, Poland;}
\affiliation{Marc Kac Complex Systems Research Centre, Jagiellonian University,
Reymonta 4, 30-059 Krak{\'o}w, Poland}
\affiliation{Copernicus Center for Interdisciplinary Studies,
30-387 Krak\'ow, Gronostajowa 3, Poland}

\author{Pawe{\l} \surname{Tambor}\email{xpt76@poczta.fm}}
\affiliation{Department of Theoretical Physics,
The John Paul II Catholic University of Lublin,
Al. Rac{\l}awickie 14, 20-950 Lublin, Poland}
\affiliation{Copernicus Center for Interdisciplinary Studies,
30-387 Krak\'ow, Gronostajowa 3, Poland}

\begin{abstract}
We discuss epistemological and methodological aspects of the Bayesian approach 
in astrophysics and cosmology. The introduction to the Bayesian framework is 
given for a further discussion concerning the Bayesian inference in physics. 
The interplay between the modern cosmology, Bayesian statistics, and philosophy 
of science is presented. We consider paradoxes of confirmation, like Goodman's 
paradox, appearing in the Bayesian theory of confirmation. As in Goodman's 
paradox the Bayesian inference is susceptible to some epistemic limitations in 
the logic of induction. However Goodman's paradox applied to cosmological 
hypotheses seems to be resolved due to the evolutionary character of cosmology 
and accumulation new empirical evidences. We argue that the Bayesian framework 
is useful in the context of falsificability of quantum cosmological models, 
as well as contemporary dark energy and dark matter problem.
\end{abstract}

\maketitle

\section{Introduction}

In everyday experience, even when we do not realize it, we use our
intuition to draw inferences.  For example when we hear doorbell we
immediately ask ourselves ``Who has come?''.  On the way to the door
we consider many different possibilities. Maybe it is our friend, who
said some days before that he visit us. Maybe it is our neighbor, who
came to say that we should lower the music or maybe somebody came to
inform us that we have won \EUR{1,000,000} in the lottery and so on
and so on. All of these possibilities are more or less probable, with
some outcomes which we cannot figure out \textit{ex ante}. But every
bit of information influences our expectations and we stick on outcome
which seems to be most probable (subjectively) to us. So we look
through the window and see a car on the street, that looks like our
friend's, we are almost sure that it is he in fact. Or if we really
play music very loudly we can expect that sooner or later somebody
will be angry.  When we do not expect somebody we know, we assume that
it may be the postman or someone who got the wrong address. In fact we
always choose the simplest case. It means that we never assume that
Superman is standing behind the door, even when there is a reason
why he should visit us.  This intuitive feeling to choose the simplest
solution is called Occam's razor \cite{Rodrigez:1999} and it
states formally
\begin{quotation}
%\textsl{
``Accept the simplest explanation that fits the data''.
\end{quotation}

In our example the data means in fact what we already know. This example 
is very easy and our faculties manage very well to solve problems of this 
kind.  However there is a vast variety of problems for which there are 
uncountable possibilities and such problems can be treated only in an 
approximated manner. On the other hand there are problems, which are too 
tedious to be solved by humans in reasonable time and we should employ to 
do job. So, can we enclose this intuitive knowledge in the form of 
mathematically defined theory and use it instead of mind? The answer 
is Yes, this exciting idea is embodied in the form of Bayesian inference 
\cite{MacKay:2003it}.

Below we introduce Bayesian inference and show how it works in practice. 
We start from general considerations which lead us to the connection with 
thermodynamics. Subsequently we show how to write down problems on computer 
with use of the Monte-Carlo approach (the Metropolis algorithm). We describe 
possible applications in the modern cosmology. There is a huge number of 
places in cosmology where Bayesian inference can often be applied. We have 
residual observations and a lot of theories. Some of them are easy, some are 
pure models, some are brilliant new ideas. This is as with our example with 
the door bell. We hear the bell and we must predict who is at the door.
Without any information we cannot predict who is ringing, because the sound 
of the bell is always the same.  However some people ring only once and some 
of them more times.  Therefore we must listen very carefully when something 
is ringing in cosmology.

In the last section we try to put Bayesian inference into a larger 
perspective: containing epistemological aspects of the method as well as 
suggested limitations.

\section{Bayesian inference and thermodynamics}

Our goal it to present a~way to choose among alternative theories taking 
into account their conformity with data. Of course these theories can 
have different basis. They can be connected with everyday experience, data 
analysis, biology, physics etc. Because we want to apply finally Bayesian
inference in physics we can restrict ourselves now, without loss of
generality, to physical theories. So let us consider some unknown
physical phenomenon and let us say we possess $K$ theories
$\{H_1,\dots,H_{\textrm{K}} \}$ that can potentially describe it. All
these theories differ from one another. Information about phenomenon
investigated is contained in the collection of experimental data $D$.
Both theories and experimental data we may consider as elements of the
same set space $\Omega$ so $\{H_1,\dots,H_{\textrm{K}} ,D\}\in
\Omega$. The set $\Omega$ together with measure $P$ and
$\sigma$-algebra $\mathcal{F}$ build probability space
$(\Omega,\mathcal{F},P)$.  In such a well defined theory of
probability, a natural concept of conditional probability occurs.  So
the probability of a given theory $H_i$, when we have data $D$, is
defined as
\begin{equation}
  P(H_i|D)= \frac{P(H_i\cap D)}{P(D)}.
  \label{posterior}
\end{equation}
This probability tells us which theory describes experimental data $D$
better and is called posterior probability. On the other hand we can
ask about probability of outcomes $D$ when theory $H_i$ is the true
one
\begin{equation}
  P(D|H_i)= \frac{P(D\cap H_i )}{P(H_i)}.
  \label{evidence}
\end{equation}
This probability tells us about different predictions $D$ from the
theory $H_i$ and is a marginal likelihood, commonly called
evidence. Because $P(H_i\cap D)=P(D\cap H_i )$ we can combine
equations (\ref{posterior}) and (\ref{evidence}) what give us
\begin{equation}
  P(H_i|D)=\frac{P(D|H_i)P(H_i)}{P(D)}.
  \label{Bayes}
\end{equation}
This equation is the famous Bayes theorem. The probability $P(H_i)$ in
this equation is called prior probability and it is in fact hard to
describe this number. It describes our initial beliefs about a given
theory. It is a human factor to choose this number and can be
non-objective. Sometimes one theory is chosen because of its
mathematical beauty although a~more probable alternative exists. The other
factor is that some theory can well describe variety of others similar
phenomena. However, if we do not have strong motivation to introduce
some initial selection of the theories, then the most natural choice
is to assume a homogeneous distribution of the prior probability
\begin{equation}
  P(H_i) = \frac{1}{K}.
\end{equation}
Then none of theories is favored. The probability $P(D)$ is a~simple 
normalization constant, what we can calculate thanks to
the normalization condition
\begin{equation}
  \sum_{i=1}^{K} P(H_i|D)= 1 ,
\end{equation}
which together with the Bayes theorem~(\ref{Bayes}) give us
\begin{equation}
  P(D)=\sum_{i=1}^{K} P(D|H_i)P(H_i).
\end{equation}

Each of the theories $\{H_1,\dots,H_{\textrm{K}} \}$ contains a number
of parameters described by the vector ${\bf \theta }_i$ for
a particular theory. The simple theories (simple
mathematically) contain in general a small number of parameters. The
main increase of number of parameters enlarges the complexity of
theories. This complexity can be in some cases accepted due to
intrinsic beauty of a~mathematical structure of theory. Nevertheless
the theory which has one factor to explain a~phenomenon is preferable
over the theory which employ many factors for description of
it. Effective theories belong to the type of simple theories.

\begin{figure}[ht!]
  \centering
  \includegraphics[width=7cm,angle=0]{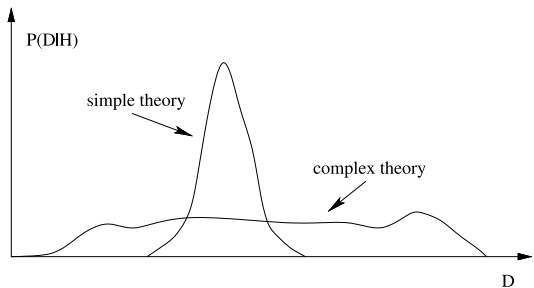}
  \caption{Evidence for simple and complicated theories}
  \label{fig:1}
\end{figure}

\begin{equation}
  P(D|H_i) = \int d {\bf \theta }_i P(D|{\bf \theta }_i,H_i)P({\bf \theta }_i |H_i).
\end{equation}

Parameters $\Theta_{i}$ of the model $H_{i}$ are elements of the set
space $\Omega$. The values of these parameters may be fixed or
significantly bounded by a~theory. But when no limits are put on these
parameters (there is no prior knowledge) the evidence is calculated as
the marginal probability integrated over all the allowed range of
values of the parameters of the model.

Now we can go back to the Bayes theorem and explain the idea of
Bayesian inference. Considering the Bayes theorems for two models $i$ and
$j$ and dividing the respective equations (\ref{Bayes}) by sides we obtain
\begin{equation}
  B_{ij} \equiv \frac{P(H_i|D)}{P(H_j|D)} = \frac{P(H_i)}{P(H_j)} \frac{P(D|H_i)}{P(D|H_j)},
\end{equation}
which is called the Bayes factor. If the priors $P(H_i)$ for all $i$  are 
equal then the Bayes factor reduces to the ratio of evidences 
($B_{ij}=P(D|H_i)/P(D|H_j)$). The values of $B_{ij}$ can be interpreted as 
follows: if $ 0 < \ln B_{ij} < 1$ then inference is inconclusive, 
if $ 1 < \ln B_{ij} < 2.5 $ we have weak, if $ 2.5 < \ln B_{ij} < 5 $
we have moderate and if $ 5 < \ln B_{ij}$ we have strong evidence in favor 
of a model indexed by $i$ over the model indexed by $j$ \cite{Trotta:2008qt}.

So the main problem is now to calculate the evidence. The direct
calculation is generally impossible.  That is the reason is to use the
Monte Carlo
methods to do it. First we introduce the parameter $\lambda$ and
redefine evidence to the form
\begin{equation}
  P(D|H_i)_{\lambda} = \int d \pi_i P^{\lambda} (D|{\bf \theta }_i,H_i)
\end{equation}
where
\begin{equation}
  d\pi_i =  d {\bf \theta }_i P({\bf \theta }_i |H_i).
\end{equation}
The $P(D|{\bf \theta }_i,H_i)$ is in fact likelihood and we denote it by $L$.  So
\begin{equation}
  \frac{d \log P(D|H_i)_{\lambda}  }{d\lambda} =
  \frac{\int d \pi_i L^{\lambda} \log L  }{\int d \pi_i L^{\lambda}} \equiv
  \langle \log L \rangle_{\lambda}
  \label{eq11}
\end{equation}
and
\begin{equation}
  P(D|H_i) = \exp\left[ \int_0^1 d\lambda \frac{d \log P(D|H_i)_{\lambda}  }{d\lambda}  \right]
  = \exp \int_0^1 d\lambda  \langle \log L \rangle_{\lambda}.
  \label{eq12}
\end{equation}

Now we can show a~connection between our approach and thermodynamics.
Introducing
\begin{eqnarray}
  E   &=& - \log{L}   \\
  1/T &=& \lambda     \\
  Z   &=& \int d \pi L^{\lambda},
\end{eqnarray}
we obtain
\begin{equation}
  Z =  \int d \pi e^{-E/T},
\end{equation}
and equation (\ref{eq11}) takes a known form
\begin{equation}
  \langle E \rangle_{T} = \frac{\int d \pi E  e^{-E/T}}{\int d \pi e^{-E/T}}.
  \label{int}
\end{equation}
This is the energy of the system in the temperature $T$. When we calculate it 
for different temperatures we can directly evaluate the integral in the 
expression (\ref{eq12}) and hence the evidence. As we see, to perform the 
Bayesian inference we have to calculate the thermodynamical 
integral~(\ref{int}). This kind of integrals can be solved analytically only 
in case of very simple systems. Numerical methods to solve this kind of 
problems are known as the Monte Carlo. It is not the subject of this paper to 
describe how they work in detail. However to make this paper self-contained 
we add a short Appendix A introducing basics of the Monte Carlo methods. 
We also present experimental demonstration of property of ergodicity which 
is important in the context of Monte Carlo simulation (see Appendix B). 
An interested reader can find more on Monte Carlo simulations e.g. 
in \cite{MacKay:2003it}.

\section{A simple example} \label{SimpleExample}

Now we have all theoretical equipment to show this approach in action. In 
this example we show how to perform the Bayesian inference in a very simple 
case. We consider a very simple kind of theories and a small sample of 
data-points to make computer computation short. We also design it for 
clarity and better understanding. However generalizations to more advanced 
problems are straightforward. In the next section we will mention how to 
apply Bayesian methods to more complicated problems.

Let us consider some experiment in which we perform measurements of 
some physical variable $y$ for six different values of parameter $x$. 
In the experiment we also measure standard error of the outcomes $y$. 
In fact we one can repeat many times measurements of $y$ for a given $x$ value.
Then one can obtain the mean values of parameter $y$ together with 
its dispersion. These data points we present in Table~\ref{tab:1}.
\begin{table}
\begin{tabular}{ccc}
\hline 
$x$  &  $y$  &  $\Delta y$   \\
\hline
1 &	1 &	7 \\
2 &	3 &	3 \\
3 &	5 &	4 \\
4 &	7 &	6 \\
5 &	10 &	3 \\
6 &	15 &	1 \\
\hline
\end{tabular}
\caption{In the table we collect the exemplary pairs $(x,y)$ together with the 
uncertainty of $y$. The uncertainty $\Delta y$ can be the result of the 
instrumental resolution.}
\label{tab:1}
\end{table}
The phenomenon which we instigate is still not undetermined, but we possess 
three polynomial models to describe them. We list these models below
\begin{eqnarray}
\textrm{model 1:} \quad y_{1}(x) &=& \alpha_0 + \alpha_1 x, \\
\textrm{model 2:} \quad y_{2}(x) &=& \alpha_0 + \alpha_1 x + \alpha_2 x^2,  \\
\textrm{model 3:} \quad y_{3}(x) &=& \alpha_0 + \alpha_2 x^2.
\end{eqnarray}
Models 1 and 3 look more simple because each is described by two parameters 
when model 2 contains three unknown parameters.

The first step of Bayesian inference is to fit these models to experimental
data. We can use for example method of least squares. We obtain
\begin{eqnarray}
\textrm{model 1:} \quad \alpha_0 &=&-2.47\pm 1.07, \alpha_1=2.66\pm0.27, \\
&& \text{SSE}=1.148, \bar{R}^2 = 0.949; \nonumber \\
\textrm{model 2:} \quad \alpha_0 &=& 0.7\pm1.02, \alpha_1=0.28\pm0.67, \alpha_2=0.34\pm0.09,\\
&& \text{SSE} = 0.571, \bar{R}^2 = 0.987; \nonumber \\
\textrm{model 3:} \quad \alpha_0 &=& 1.10\pm0.33, \alpha_2=0.38\pm0.02, \\
&& \text{SSE}= 0.509, \bar{R}^2 = 0.990. \nonumber
\end{eqnarray}

The important ingredient in the Bayesian inference is a choice of priors. 
It is the choice of the intervals and probability distribution for parameters.
The parameter intervals should be specified, because we must perform the 
integration (look for the solution) in a finite parameter space.
Standard errors of the parameters give us intervals necessary for 
Bayesian inference. Of course, the different choices of the parameter 
intervals can lead to the different values of the posterior probabilities. 
In our case, we choose the intervals as the $1$-$\sigma$ confidence 
interval for the parameters. Namely, $[\alpha-\sigma, \alpha+\sigma]$, where 
$\alpha$ is the best fit value and $\sigma$ is the standard error. 
Next we assume that outcomes are from the Gaussian distribution and the 
likelihood function has then a form
\begin{equation}
L \propto \exp\left[ -\frac{1}{2}  \sum_{i=1}^{N}  \frac{(y(x_i)-y_i )^2}{\sigma_i^2} \right].
\end{equation}

\begin{figure}[ht!]
\centering
\includegraphics[width=7cm,angle=0]{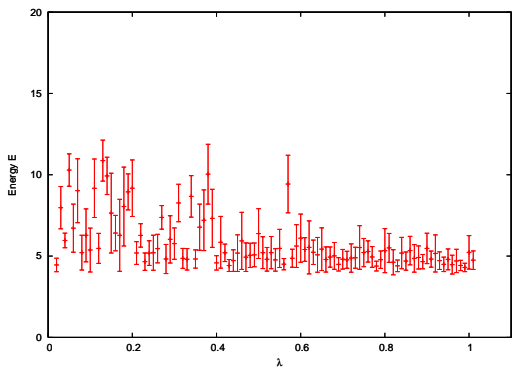}
\caption{$\langle E \rangle_{\lambda}$ dependence for the first model.}
\label{1}
\end{figure}

\begin{figure}[ht!]
\centering
\includegraphics[width=7cm,angle=0]{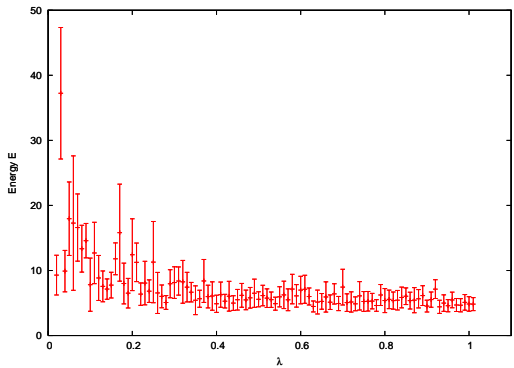}
\caption{$\langle E \rangle_{\lambda}$ dependence for the second model.}
\label{2}
\end{figure}

\begin{figure}[ht!]
\centering
\includegraphics[width=7cm,angle=0]{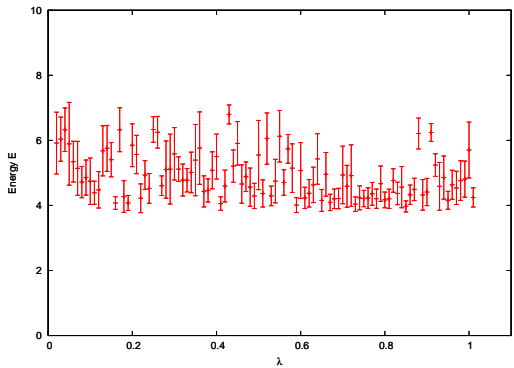}
\caption{$\langle E \rangle_{\lambda}$ dependence for the third model.}
\label{3}
\end{figure}

Now we apply theory described in the previous section. With use of the 
Metropolis algorithm we calculate energies $\langle E \rangle_{\lambda}$ for 
models considered. We perform calculations for the values of $\lambda$ from 
the range $(0,1)$ as it is necessary to calculate the integral in
equation~(\ref{eq12}). We show these results in Fig.~\ref{1},~\ref{2},~\ref{3}.

Now with use of this data we can perform integrals in the form
\begin{equation}
I_i = \int_0^1 d\lambda \langle E \rangle_{\lambda}.
\end{equation}
From equation~(\ref{eq12}) we see that
\begin{equation}
P(D|H_i) = \exp \int_0^1 d\lambda  \langle \log L \rangle_{\lambda}  = e^{-I_i}
\end{equation}
Obtained values for three models considered are
\begin{equation}
I_1 = 5.8, \ I_2 = 7.6, \ I_3 = 4.8.
\end{equation}
Now we can directly calculate Bayes factors
\begin{eqnarray}
B_{12} &=& 6.06, \ \ \ln B_{12} = 1.80, \\
B_{32} &=& 16.67, \ \ \ln B_{32} = 2.81,    \\
B_{31} &=& 2.78, \ \ \ln B_{31} = 1.01.
\end{eqnarray}
Based on the obtained values of $\ln B_{ij}$ one can conclude: 
\begin{itemize} 
\item  Because $1<\ln B_{12}<2.5$, we have weak evidence that the linear model 
is favored over the three parametric quadratic model.

\item  Because $2.5 <\ln B_{32}< 5$, the two parametric quadratic model 
is moderately preferred over the the three parametric quadratic model.

\item  Because $\ln B_{31} \approx 1$, there is no preferences between the
 two parametric models, linear and quadratic.
\end{itemize}

We can also directly calculate posterior probabilities
\begin{eqnarray}
P(H_1|D) &=& 0.26,  \\
P(H_2|D) &=& 0.04,  \\
P(H_3|D) &=& 0.70.
\end{eqnarray}
What we see is that the posterior probability suggest that the last model 
explains the experimental data in the best way. Even that it may look more 
complicated than linear model number one. The second model can be discarded. 
This model contains remaining models inside and naturally fits better to the 
experimental data because has more degrees of freedom. But comparing to the 
other models he is too complicated and not necessarily properly explains 
experimental data. We see that however first and last model possess the same 
number of degrees of freedom the last function seems to explain in the better 
way the nature of the investigated physical process. On the other hand, from 
the  Bayes factor criterion, one can not seen such a preference. It is because 
the borders in the Bayes factor criterion were established empirically in the 
very careful manner.

The example presented here illustrates that the different criteria give us 
stronger or weaker conclusions. In particular, note that posterior probability 
criterion favors the model $H_3$ over the model $H_1$. On the other hand the 
Bayes factor criterion indicates that there is no preference between the 
models $H_3$ and $H_1$. Therefore, this simple example gives us here the 
first lesson to be very careful interpreting statistical inference results.

The second lesson is to be careful about results of estimations. 
The results presented above, base on the $1$-$\sigma$ interval choice of the 
parameters intervals. In case of the $3$-$\sigma$ intervals the resulting 
Bayes factors are the following
\begin{eqnarray}
\ln B_{12}  &=& 0.4,  \\
\ln B_{23}  &=& 1.5,  \\
\ln B_{13}  &=& 1.9.
\end{eqnarray}
Therefore, there the first model is weakly favored with respect
to the third one. Also the second model is weakly favored
with respect to the third one. The first and the second models are
indistinguishable. The $3$-$\sigma$ results differ from these
performed previously.

The comparison between the $1$-$\sigma$ and $3$-$\sigma$ examples
was performed to show explicitly the sensitivity of the Bayesian
inference on the parameter intervals. This dependence on the
priors becomes significant in case of the \emph{weak data}, e.g. for
the small sample size or significant error. For the \emph{strong data}
the impact from the priors becomes irrelevant. In the case considered here,
the data sample is small and the significant errors are present.
Because of this, the inference is sensitive on the choice of the
priors as observed.

\section{Applications in cosmology}

In this section we present some applications of Bayesian inference in 
modern cosmology. The Bayesian methods have been introduced to cosmology 
relatively late. The proliferation of dark energy models forced using the 
formal methods of model selection. Starting from the pioneering work of 
M.~V. John and J.~V. Narlikar \cite{John:2002gg} the Bayesian methods have 
become popular \cite{Trotta:2008qt}. The reason is that cosmology really needs 
these methods, because our knowledge about the Universe is still very limited. 
We have a lot of theories about different stages of the Universe but only 
a modest number of observations to verify them. So it is the right place 
for the Bayesian methods.

\subsection{Dark energy and dark matter}

The first case that we would like to talk about is connected with very
mysterious behavior of the Universe, discovered in the end of the last decade. 
Namely, observations of the distant type Ia supernovae (SNIa) indicated that 
the Universe expansion accelerates \cite{Riess:1998cb}. This discovery was
apparently in the conflict with other observations and beliefs that
Universe in filled with normal matter like dust, stars, planets etc.
Given this assumption the Universe always decelerates. So what is
happening in the Universe? Why did it start to accelerate after
a previous phase of deceleration? What is the mysterious component of
the Universe that we call dark energy?  There is presently an enormous
number of possible answers for this question. The presence of the
cosmological constant $\Lambda$, phantom fluid, modified gravity,
field of quintessence, quantum gravitational effects, brane world
models, vacuum energy and so on and so on. The number of possible
solutions is really impressive. So, which one of them is the real
solution? Which one describes Nature in the right way? Or maybe none
of them, maybe we still must be looking for new models. Let us neglect
the possibility of building new models to explain acceleration of the
Universe and limit ourselves only to some already proposed solutions. 
This is precisely what we can do thanks to Bayesian inference. Due to 
Bayesian methods one can obtain a ranking of accelerating models 
\cite{Szydlowski:2006ay,Kurek:2007tb}. The analysis show that standard 
cosmological model, the so-called $\Lambda$CDM model, is on the top off all 
theoretical propositions \cite{Kurek:2007tb}.

The assumed model of the Universe is the Friedmann-Robertson-Walker 
model filled with dust matter, dark and baryonic, and dark energy. 
In Bayesian estimation of cosmological model parameters for these components 
we start from the cosmography which determines the luminosity function as the 
function of density parameters. The density parameters define the 
fraction of energy in the total energy budget of the universe. They are 
dimensionless and their sum is equal one. The additional parameter 
is the Hubble constant which describes the rate of expansion of the 
current Universe. While the kinematic part of the cosmology is controlled 
by cosmography basing on analysis of trajectories of photons, the 
observations of CMB control the dependence of perturbation on time. 

Among many theoretical propositions the simplest candidate for dark energy 
is the time-independent cosmological constant. The other propositions:
the Chaplygin gas, models with varying coefficient of the equation of state,
quintessence, phantoms, etc. These models are fitted using the astronomical 
data: the measurement of absolute magnitudes of high redshift type Ia 
supernovae, cosmic microwave background (CMB) radiation, the measurement 
of baryon acoustic peak (BAO) in galaxies correlation function, gas mass 
fraction value in galaxy clusters, gamma-ray bursts (GRB) which are at 
higher redshift than SNIa.

In selection of cosmological models with different forms of dark energy 
different criteria are used. The simplest criteria are the information 
criteria AIC, BIC and Bayes factor, posterior probability. 
These methods allow us pick out the best model in the light of data 
at our disposal.

\subsection{Can we distinguish quantum gravitational effects from
  observations?}

This intriguing question corresponds to the presence of possible
observational phenomena of quantum gravity.  The quantum gravitational
effects are predicted to be very small and unreachable by the present
and any future generations of accelerators. However the quantum
gravitational effect can survive as the relict from the very early
Universe in which quantum gravitational effects have been
dominant. These effects can influence the spectrum of inflationary
perturbations \cite{Danielsson:2002kx,Mielczarek:2008pf}.  These
primordial fluctuations then lead to the fluctuations of matter and
finally to the structures formation in the Universe. So can we deduce
some information about quantum gravity from observations of microwave
background radiation and large scale structures? At first glance it
can sound strange because quantum gravity describes microscopic
property of gravitational field at a very deep level.  However the
same effects were very important in the early universe and could
influence its global properties.  Is here a place for Bayesian
inference? The answer is \emph{Yes}. We have now a lot of predictions from
quantum theories of gravity like Loop Quantum Gravity
\cite{Ashtekar:2004eh} and still a very limited number of observations
in the region where they can be important, e.g. non-gaussianity in
primordial fluctuation in CMB or the primordial gravitational waves
spectrum.

The two examples presented in this section are very important but they
are not the only ones. There is a lot of other places in cosmology
where Bayesian inference is and should be applied.  For example, we
still do not know what the dark matter (the second dominant component
of the Universe) is. It may be an axion, higisno, gluino other super
particles \cite{Jungman:1995df} or just neutrinos etc. It is an ideal
place for the Bayesian inference to point out the best candidate.
Unfortunately still too little is known from observations.

\section{Some general epistemic remarks on Bayesian inference}

It is an obvious fact that Bayes' theorem is a theorem. Nevertheless,
it should be explicitly explained, that to be a Bayesian is more than
\textit{to know and use the theorem}. The Bayesian theory of confirmation
achieves a great success in such areas of human activity as physics,
biology, medicine, cognitive science (decision theory) on the one
hand, and encounters serious limitations of the analyzed method on the
other hand. A radical Bayesian would probably say that it can always
be applied: we always have to do with a joint distribution and
a numerical representation of belief is always possible. However,
a Bayesian epistemology could be treated not only as competing with
other methods of confirmation, but as a way of making a specific (not
exclusive) model of beliefs and related evidences, which can be easily
elaborated and understood.

In the Bayesian approach probability is attributed to hypotheses that
are being confirmed. This confirmation can be interpreted both
qualitatively and quantitatively, since inference is based on
empirical data and there are relations between hypotheses, theories
and observations to be explicated. It is indeed a crucial point in
understanding Bayesian inference -- the meaning which is ascribed to
probability. Using probabilistic methods one can measure two things:
how often a specific event occurs and how strong evidence (confirming
our beliefs) is.

Let us generally state that Bayesianism can be treated as an epistemic
theory which examines the relation between beliefs and empirical
evidence as to measure the strength of the beliefs. As it has been
shown the most important concept used to gain that goal is the notion
of conditional probability \cite{Strevens:2006}. Using the Bayesian
inference we not only measure the strength of beliefs but also propose
the method for rational estimating a change of the beliefs under the
influence of a new evidence.

Sometimes, among scientists and philosophers of science, there is
a bit of hesitation about exclusiveness of such an approach. S. Okasha
in his study on van Frassen's conception of induction wrote
\cite[p. 693]{Okasha:2000}:
\begin{quotation}
  He accepts the Bayesian representation of opinion in terms of
  degrees-of-belief, and he agrees that synchronic  probabilistic
  coherence is a necessary condition of rationality. However, he does
  not accept the Bayesian thesis that conditionalization is the only
  rational way to respond to new evidence; though he allows that it is
  a rational way.
\end{quotation}

It can be said in that sense that Bayesianism offers a solution to old
problems of induction. We have got an approximately coherent and
reasonable model for probability corrections. Of course, it is
possible if having initial probability and evidence (priors). The
proposed solution has its weakness: its method often tells nothing how
to estimate these probabilities.

The crucial point in Bayesian inference lies in the fact that it is
able to deal with the problem, only if we manage the input of some
probabilities. It makes sense, since we never start reasoning with
absolutely no knowledge. The result we achieve -- $P(H|E)$ -- are always
``conditional'': it reveals a property of $H$ which is not objective,
but related to $E$ and certain \textit{knowledge}, called
\textit{background knowledge}. A more subtle epistemic analysis can be
carried out with reference to the types of background
knowledge. P. Wang studied the problem of underlying knowledge and
discerned two types of conditions which influence the evaluation of
probability distribution function \cite[pp. 98-99]{Wang:2004} in the
formulas, as follows:
\begin{itemize}
\item explicit conditions
  \[
  P(H|K_{1}) = P(H|E \wedge K_{0}) = \frac{P(E|H \wedge
    K_{0})P(H|K_{0})}{P(E|K_{0})}
  \]
  \begin{itemize}
  \item $E$ is a binary proposition,
  \item belongs to proposition space (then we can define $P_{0}(E)$),
  \item $P_{0}(E) > 0$;
  \end{itemize}
\item implicit conditions
  \[
  P_{K_{1}}(H) = P_{K_{0}}(H|E) = \frac{P_{K_{0}}(E|H) P_{K_{0}}(H)}
  {P_{K_{0}}(E)}
  \]
  \begin{itemize}
  \item non-binary propositions allowed,
  \item there may be statements outside the proposition space,
  \item ``Even if a proposition is assigned to a prior probability of
    zero according to one knowledge source, it is possible for the
    proposition to be assigned a non-zero probability according to
    another knowledge source''.
  \end{itemize}
\end{itemize}

All these discernments are not trivial, since we have to answer the
question: \textit{whether all the background knowledge can be
  probabilistic-valued?}. This is one of the most important epistemic
questions of Bayesian Theory of Confirmation, beside the others:
\begin{itemize}
\item \textit{Are there degrees of belief?} The answer, maybe, lies in
  an attempt to distinguish `rational' degrees of belief from belief
  in general. Are the corrections in probability, gained in the
  Bayesian procedure, just new probabilistic information or do they
  deliver \textit{new reason to believe} that the proposition
  considered is true?
\item When are we actually updating our belief and when there is just
  a revision of probability (known problem of old evidence)?
\end{itemize}

While elaborating empirical data in cosmology, one can use classical or
Bayesian statistics. In a classical approach we rely on the classical
definition of likelihood, but doing the same \textit{Bayesian way} we 
are dealing with \textit{a priori} and \textit{a posteriori} probability, 
respectively. There is an opinion among some cosmologists that the standard
approach (\textit{the standard maximum-likelihood technique}) is satisfactory
indeed, which is reasonable if there is a single model under consideration. 
When we have several competitive hypotheses, Bayesian statistics would be 
a better choice.

There are two groups among Bayesians who differ from each other with
respect to criteria used in choosing of priors: objective Bayesians:
E.~T. Jaynes \cite{Jaynes:2003}, H. Jeffreys \cite{Jeffreys:1961},
R.~D. Rosenkrantz \cite{Rosenkrantz:1977}) and subjective Bayesians:
B.De~Finetti \cite{DeFinetti:1974}, C. Howson and P. Urbach \cite{Howson:1989}.

It is not only the problem (or problems) of induction, that Bayesian
inference tries to deal with, but also the problem of finding
a justification of induction inference itself, which can be explicated
in several schemas:
\begin{itemize}
\item Inductive Generalization \\
  Nobody denies that a finite number of experimental data cannot
  deliver an exhaustive proof to a universal statement but, according
  to induction, empirical evidence confirms generalization
  \cite{Carnap:1950lf,Reichenbach:1949}. An example of that is
  enumerative induction which principle explicates, as follow:
  \begin{quote}
    Several crows are black. \\
    Therefore, all crows are black.
  \end{quote}
  \begin{quote}
    Every load added so far has not damaged this truck. \\
    Therefore, the next piece of load will not damage this truck.
  \end{quote}

\item Hypothetical Induction \\
  It occurs when some hypothesis deductively entails the evidence.
  \begin{quote}
    An evidence confirms hypothesis, if the evidence is a logical
    consequence of the hypothesis.
  \end{quote}
  In the case of existing multiple competing hypotheses one can try to
  show that the falsity of the hypothesis entails the falsity of the
  evidence or use additional criteria of hypotheses' selection, like
  simplicity or inference to the best explanation. However, these
  proposals create \textit{market of hypotheses} (for example
    cosmological models) with rules for successful selection but in
  fact they cannot give any rational explanation to the evidence.  It
  may be a truism, but the difference between explanation and
  confirmation should be treated with special care. The more so
  because there is not a unity among Bayesians concerning
  representation of the degree to which evidence supports an
  hypothesis. The most popular are three options:
  \begin{itemize}
  \item a difference measure: $P(H|E) - P(H)$;
  \item a normalized difference measure: $P(H|E) - P(H| \neg E)$;
  \item a likelihood measure: $\frac{P(H|E)[1-P(H)]}{[1-P(H|E)] P(H)}$
  \end{itemize}
\end{itemize}

The inductive generalization, which has the simple pattern:
extrapolation from particular data do general conclusions, suffers
several problems called paradoxes of confirmation. Goodman's paradox,
know in the literature as problem of ``grue'', is particularly
interesting \cite{Goodman:1955}. Especially the question of its
counterpart in the field of cosmology. In a traditional version:
\begin{itemize}
\item We have two hypotheses: (1) all emeralds are green and (2) all
  emeralds are grue (green if examined until some time $t$ and blue
  otherwise).
\item Evidence: \textit{found emerald is green} confirms both: (1) and
  (2).
\end{itemize}
Any satisfying resolutions to the paradox propose additional
assumptions; for example pointing out on ``green'' as a natural kind
term instead of ``grue''.

In a search for possible cosmological version of the paradox we can compare 
for example two related models the cold dark matter cosmological model (CDM 
model) and the Lambda cold dark matter cosmological model with the positive 
cosmological constant term (LCDM model). The latter seems to be the simplest 
candidate for the dark energy description. The Bayesian method of confirmation 
dedicated to select between these two models reveals a quite opposite verdict 
while used in the 90s and currently. Using the sample of Perlmutter et al. 
\cite{Perlmutter:1997zf} there is not enough information to distinguish these 
models. The extended sample with additional 42 high $z$ SNIa 
\cite{Perlmutter:1998np} gives a weak evidence to favor the LCDM model over 
the CDM one. However, in our opinion it is a misunderstanding to treat this 
study case as a paradox in Goodman's sense. It becomes obvious, because when 
new observational data confirm better the LCDM model in comparison with the 
CDM model, the latter simply disappears out the stage. The paradox of 
confirmation would occur when related to a certain family of models there will 
be the same degree of confirmation (the same time and evidence) assigned to 
hypotheses differing from each other for example with regard to foreseeable 
future scenarios of Universe evolution\footnote{Historically these two 
hypotheses have never coexisted in the same time. Until the late 90s the 
hypothesis of the CDM model was accepted by cosmologists, but SNIa 
observations made that the new hypothesis of the LCDM model was necessary 
to be formulated.}.

To illustrate this situation let us consider two hypotheses

\begin{enumerate}
\item The Universe decelerates.
\item The Universe decelerates until some time $t$ and
  accelerates afterward.
\end{enumerate}

The CDM model is valid with the first hypothesis and the LCDM model is
in agreement with the second one. From the 60's it was known that the
Universe is expanding with the decelerating rate. So we have
a~paradox here. However the evidences of accelerating Universe due to
SNIa data falsified the first model. And the paradox is naturally
solved. This example teaches us that paradoxes of Goodman's type (in
the logic of induction) are common in evolutionary sciences but they
are not dangers because we hope that new evidences (which appear due
to science development) we discriminate between two hypotheses.

In Goodman's paradox there is only one kind of evidence; we need to
draw an emerald and check its color. In the case of cosmological
hypotheses we are not left with only one evidence. A~new evidence
appears and resolves the paradox in favor of one of the
hypothesis. It comes from new observations. We know that this evidence
will appear eventually because we, scientists look for it. The reason
that there is no paradox after a~new evidence appears, is that one
hypotheses is falsified (the CDM model) and only one hypothesis (the
LCDM model) becomes in agreement with this new evidence.

It is often said that a scientific theoretical research means
achieving two specific goals: (1) finding a model which approximates
a phenomenon best and (2) constructing a hypothesis that offers best
prediction. It is a good example to show how in this context two
criteria of model selection are being compared: the Akaike information
criterion (AIC) and Bayesian information criterion (BIC)
\cite{Liddle:2007ez}. Although these model comparison methods are put
together as competitors, they in fact try to ask different questions
\cite{Szydlowski:2008by}. The AIC estimates predictive power of an
elaborated hypothesis, while the BIC -- goodness-of-fitting
\cite{Sober:2002}. M. Forster and E. Sober have explained this nuance
with respect to the fitting problem \cite[pp. 5-9]{Forster:1994}:
\begin{quotation}
  Even though a hypothesis with more adjustable parameters would fit the data 
better, scientists seem to be willing to sacrifice goodness-of-fit if there is 
a compensating gain in simplicity.(\ldots)\\
  Since we assume that observation is subject to error, it is
  overwhelmingly probable that the data we obtain will not fall
  exactly on that true curve.(\dots) Since the data points do not fall
  exactly on the true curve, such a best-fitting curve will be
  \textit{false}. If we think of the true curve as the `signal' and
  the deviation from the true curve generated by errors of observation
  as `noise', then fitting the data perfectly involves confusing the
  noise with the signal. It is overwhelmingly probable that any curve
  that fits the data perfectly is false.
\end{quotation}

The general comments of this section can be summed up by a statement
that Bayesian inference is a method dedicated to specific goals in
scientific practice \cite{Linder:2007fv}.  With respect to cosmology,
the mentioned LCDM--CDM models comparison reveals in Bayesian
inference context another problem. It strictly concerns currently
changing concept of the model in physics \cite{Morrison:2005}. At
present there is a special emphasis placed on effectiveness and
mediating function of models in physics. This
status of scientific models is determined by the way they are
designed: they are not simply derived from the underlying theory, nor
fixed by the evidence only. Their ``nature'' is determined by
a mediating role (between a theory and phenomena). Morrison states, as
follows \cite[p. 67]{Morrison:1998}:
\begin{quotation}
  Although they are designed for a specific purpose these models have
  an autonomous role to play in supplying information, information
  that goes beyond what we are able to derive from the data/theory
  combination alone.
\end{quotation}

In cosmology built on general relativity, the solutions of Einstein equations 
can be treated as the geometrical models of the Universe. A construction
of a model starts from assuming specific idealizations (symmetries, etc). 
It means in a practice that we reduce degrees of freedom (all apart gravitational 
ones are neglected). For example assumption of spacial homogeneity means that 
the Einstein equations which constitute the system of non-linear partial 
differential equation system reduce to the ordinary differential equation 
system in the cosmological time. It is said that those
formulations of scientific laws are certain approximations of the
investigated phenomena. There has been recently quite an important and
interesting discussion about validity of application the Bayesian
inference to idealization itself \cite{Shaffer:2001,Jones:2007}. The
problem concerns idealized hypotheses and a question of assignment
probability to them, since they can be treated as
counterfactuals. What is a posterior probability of the ideal gas law
or the law of motion for simple pendulum? Jones showed that solution
lies exactly in the understanding of the procedure of elaborating
a model. If we treat the model idealizations not as a result of
abstraction but as a distortion, the methodological consequences may
exclude for Bayesian inference\cite[p. 3]{Jones:2007}:
\begin{quotation}
  Given that most scientific hypotheses are idealized in some way, 
Bayesianism seems to entail that most scientific hypotheses cannot be confirmed. \\
  Bayesians thus confront an apparent trilemma: either develop
  a coherent proposal for to assign prior probabilities to
  counterfactuals; or embrace the counterintuitive result that
  idealized hypotheses cannot be confirmed; or reject Bayesianism.
\end{quotation}

The general Bayesian conception of empirical evidence can be put into
three main statements/consequences:
\begin{itemize}
\item Less probable evidence delivers best confirmation to hypothesis;
\item Evidence confirms better those hypothesis in context of which it
  is more probable.
\item If the hypothesis' probability is very little, it can be
  confirmed only by very strong evidence.
\end{itemize}

\section{Summary}

In this paper we have presented basics of Bayesian inference and showed how
to use it in practice. We have introduced some mathematical background and
formulated a problem in the similarity with thermodynamics. As a case study 
we choose three simple models. Then the known
Monte Carlo methods and Metropolis algorithm were used to select the best model 
in the light of data.
The general remark which can be derived from these considerations is to be careful 
in evaluation of the models in the light of the data and in using the complementary 
indicators.

The Bayesian methods started to be popular due to new discoveries in cosmology 
at the beginning of XXI century. We presented the areas of cosmology where 
Bayesian inference has been applied, namely problem of dark energy, dark 
matter, and testing quantum effects by astronomical data. 

Subsequently we have studied epistemological aspects of the Bayesian 
confirmation theory in the context of problems of modern cosmology where the 
Bayesian approach offers not only the estimation of model parameters from the 
observational data but also methods of the comparison of models (selection). 
We have demonstrated that the Bayesian inference is based on some assumptions 
of philosophical character. The philosophical issues of inference in context 
of cosmological models on the example of models without and with the dark 
energy component (the cosmological constant) are discussed.

We pointed out that Goodman's famous paradox does not appear in the
cosmology reconstructed using Bayesian methodology. The reason for this 
we are looking for new evidences which
falsify one hypothesis such that only one hypothesis becomes in
agreement with observational data. Note that the Bayesian framework
enable us to test and select between competing hypotheses so one can
construct the ranking of cosmological models explaining acceleration of
the current Universe. Therefore, we obtain the best model favored by data.

\appendix
\section{Monte Carlo method, Markov chain and Metropolis algorithm}

In this appendix we show how to compute thermodynamical integral
(\ref{int}) with use of the Monte Carlo simulations. Our short
introduction to this subject is based partially on this made in
\cite{Huang:2001}. Let us consider state of the system
labeled by $\Gamma$ and corresponding energy $E(\Gamma)$. Our task is
to compute integral
\begin{equation}
  \langle E \rangle_{T} = \frac{\int d \pi E(\Gamma)  e^{-E(\Gamma)/T}}{\int d \pi e^{-E(\Gamma)/T}}.
  \label{intApp}
\end{equation}
where integration is performed over all available states
$\Gamma$. Since in numerical computations we always discretize the
system, integration $\int d \pi$ is replaced by the summation.  Our
task now is to write a program which generates states $\Gamma$ from
the canonical ensemble given with the probability $
e^{-E(\Gamma)/T}$. The crucial observation is that we do not have to
generate all possible states to calculate (\ref{intApp}). The
main contribution to their value comes from the equilibrium
states. Therefore the idea is to find these equilibrium states and
average over them. Starting from some arbitrary initial state
we create a sequence of states
\begin{equation}
  \underbrace{\Gamma'_{1}\rightarrow \cdots \rightarrow \Gamma'_{K'}}_{\textrm{non-equilibrium}} \rightarrow
  \underbrace{\Gamma_{1}\rightarrow \cdots \rightarrow \Gamma_{K}}_{\textrm{equilibrium}}
\end{equation}
finally finding ensemble of equilibrium states. Then one can calculate
\begin{equation}
  \langle E \rangle_{T} \simeq \frac{1}{K}\sum_{i=1}^{K} E(\Gamma_{i}).
\end{equation}
In order to find equilibrium states the Markov chain method can be
applied. We consider sequence of transitions
$\Gamma_{i}\rightarrow\Gamma_{i+1}$ with probability
$P(\Gamma_{i}|\Gamma_{i+1})$. Moreover we assume
\begin{eqnarray}
  P(\Gamma_{i}|\Gamma_{i+1}) &\geq& 0, \\
  \sum_{\Gamma_{i+1}} P(\Gamma_{i}|\Gamma_{i+1}) &=& 1, \\
  e^{-E(\Gamma_{i})/T}P(\Gamma_{i}|\Gamma_{i+1}) &=& e^{-E(\Gamma_{i+1})/T}P(\Gamma_{i+1}|\Gamma_{i}).
\end{eqnarray}
Practical realization of the above conditions is given by
\emph{Metropolis algorithm}. Namely it states:
\begin{itemize}
\item Take initial state $\Gamma_i$.

\item Make some move to neighboring state $\Gamma_{i+1}$.

\item If $E(\Gamma_{i+1})< E(\Gamma_{i})$, accept the change.

\item If $E(\Gamma_{i+1})> E(\Gamma_{i})$, accept the change
  conditionally with the probability
  $e^{-[E(\Gamma_{i+1})-E(\Gamma_{i})]/T}$.
\end{itemize}

\begin{figure}[ht!]
  \centering $\begin{array}{cc}
    \includegraphics[width=6cm]{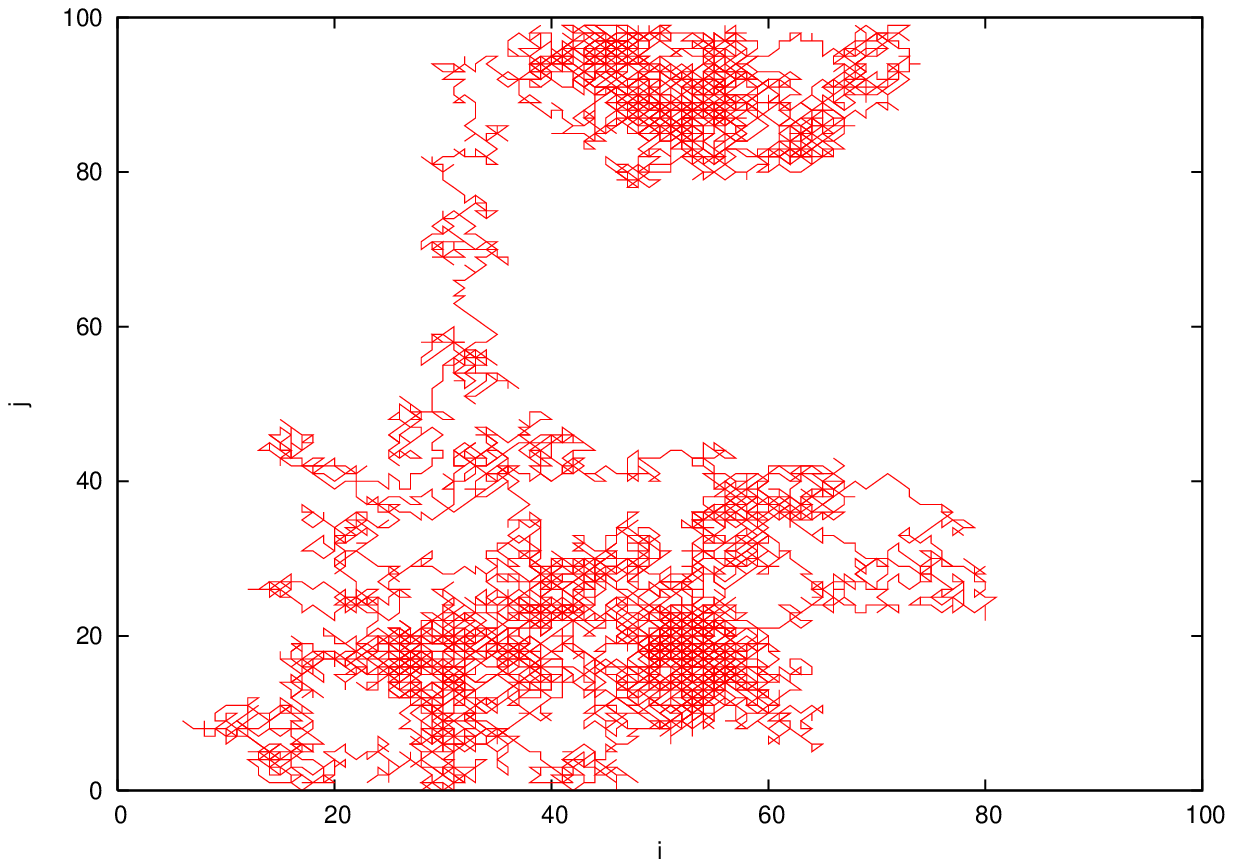}  &  \includegraphics[width=6cm]{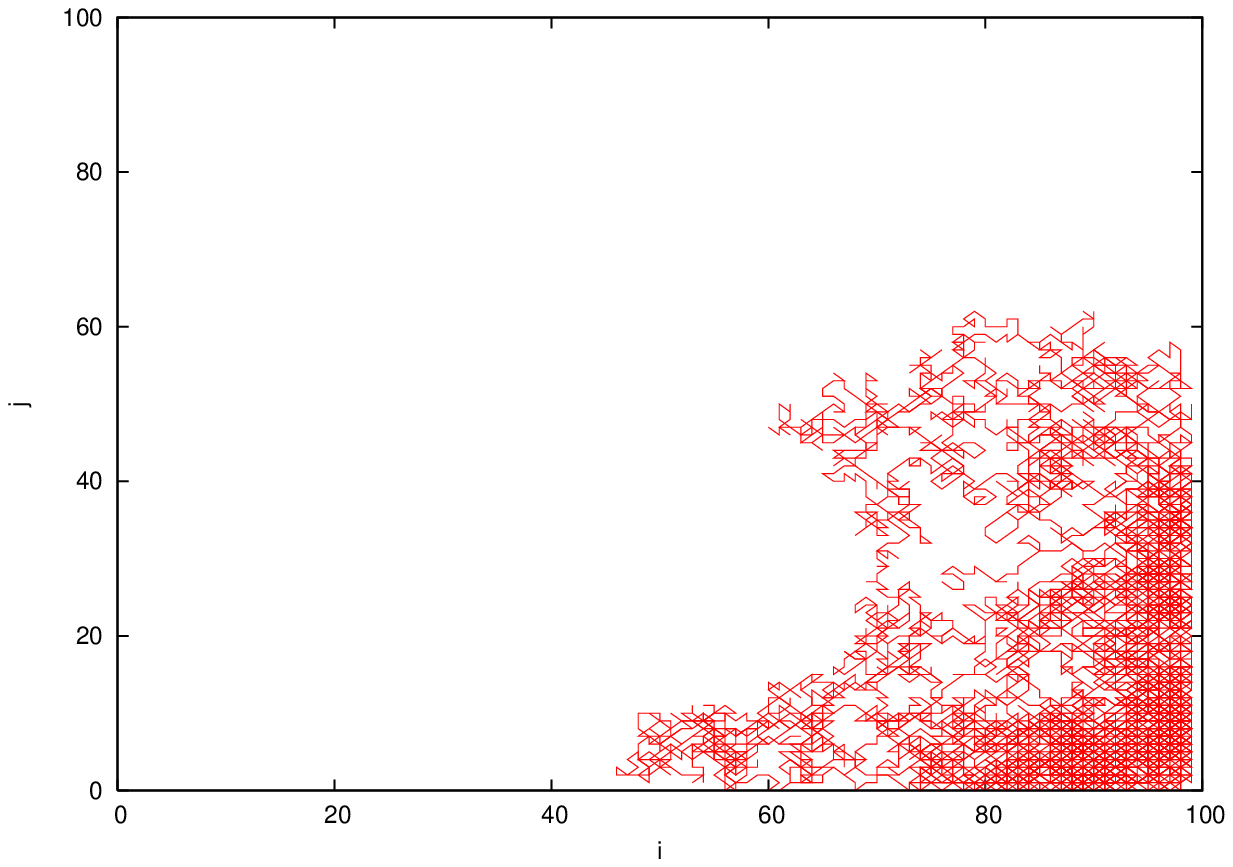}   \\
    \includegraphics[width=6cm]{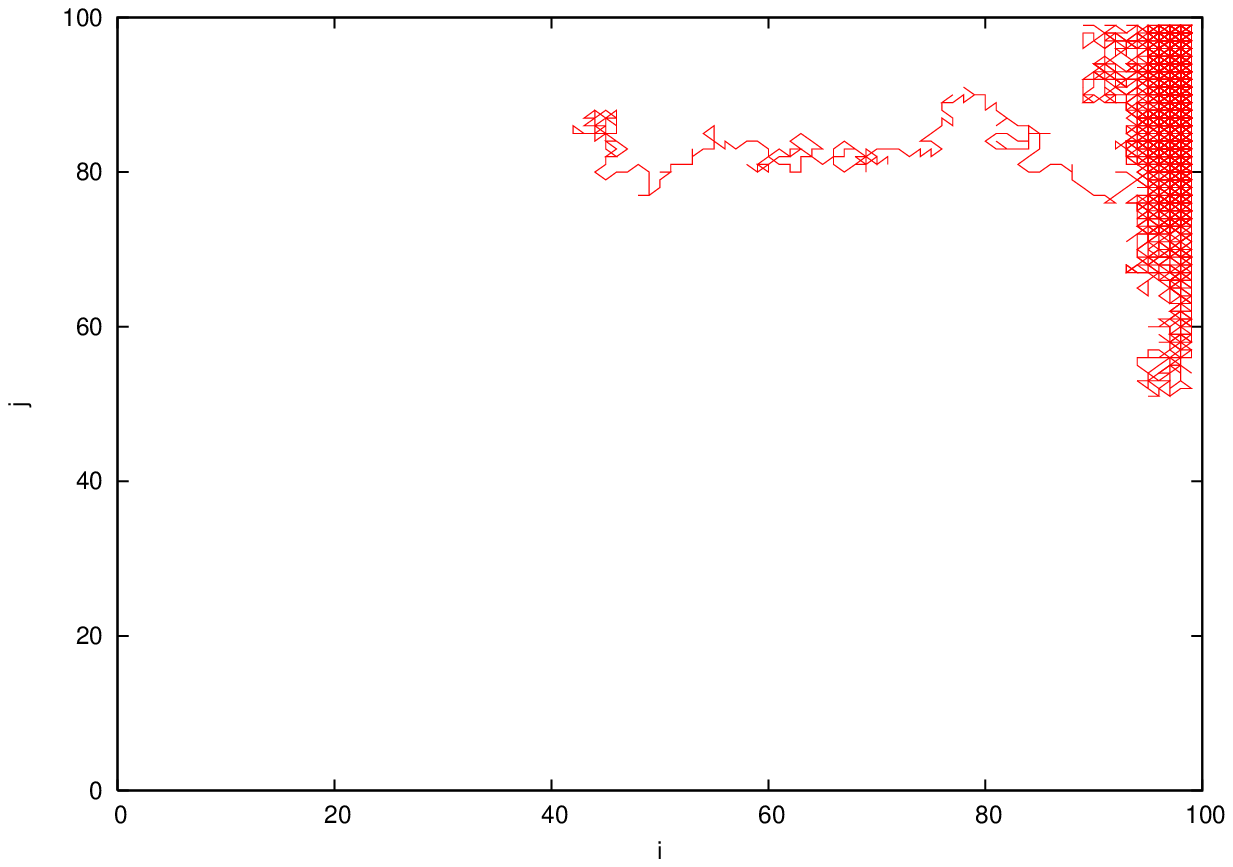}  &  \includegraphics[width=6cm]{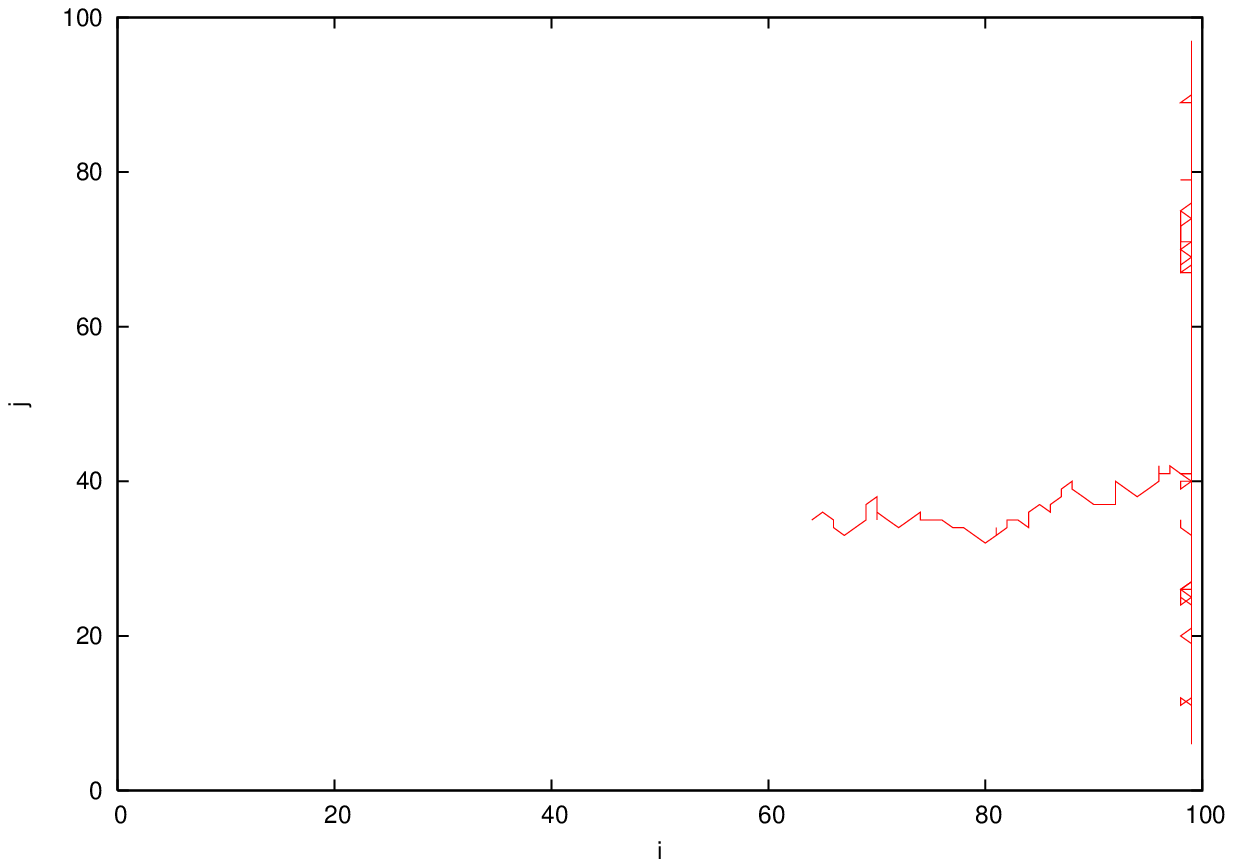}
  \end{array}
  $\caption{ {\bf Top left}: $\lambda=0.1$.  {\bf Top right}:
    $\lambda=1$.  {\bf Bottom left}: $\lambda=10$.  {\bf Bottom
      right}: $\lambda=100$.}
  \label{fig:1L}
\end{figure}

All computations performed in Sec.~\ref{SimpleExample} have been done
applying this simple set of rules. As an example we show here the
Markov chains in the parameter space for the models considered
there. In Fig.~\ref{fig:1L} we show sequence of moves for the first
model considered in Sec.~\ref{SimpleExample}.  We assume values of the
parameter $\lambda = 0.1,1,10,100$. Since $1/\lambda=T$ the higher
value of $\lambda$ corresponds to lower temperatures. We investigate
here a broad range in $\lambda$, however for the calculations of the
evidence only values of $\lambda \in [0,1]$ are required.

\section{Numerical demonstration of property of \emph{ergodicity}}

\begin{figure}[ht!]
\centering
$\begin{array}{cc}
\includegraphics[width=6cm]{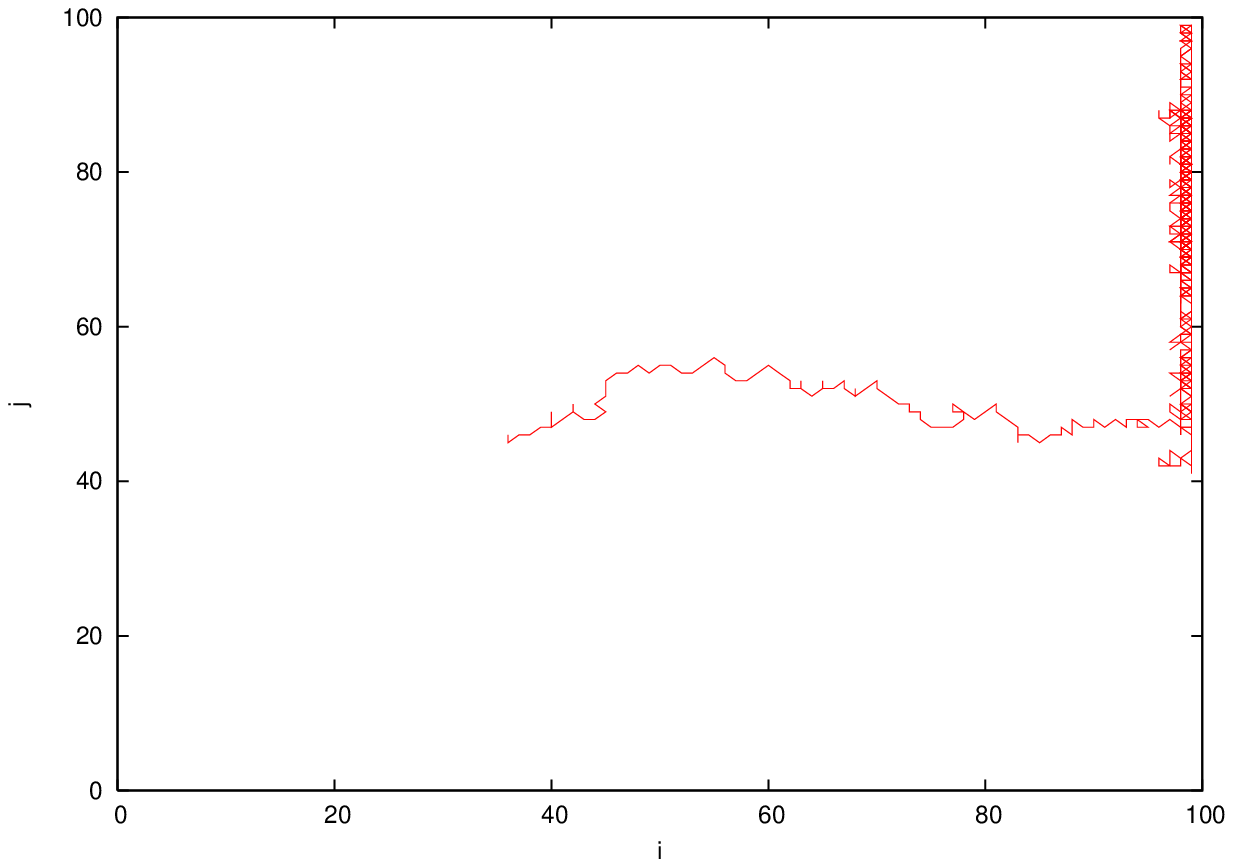}  &  \includegraphics[width=6cm]{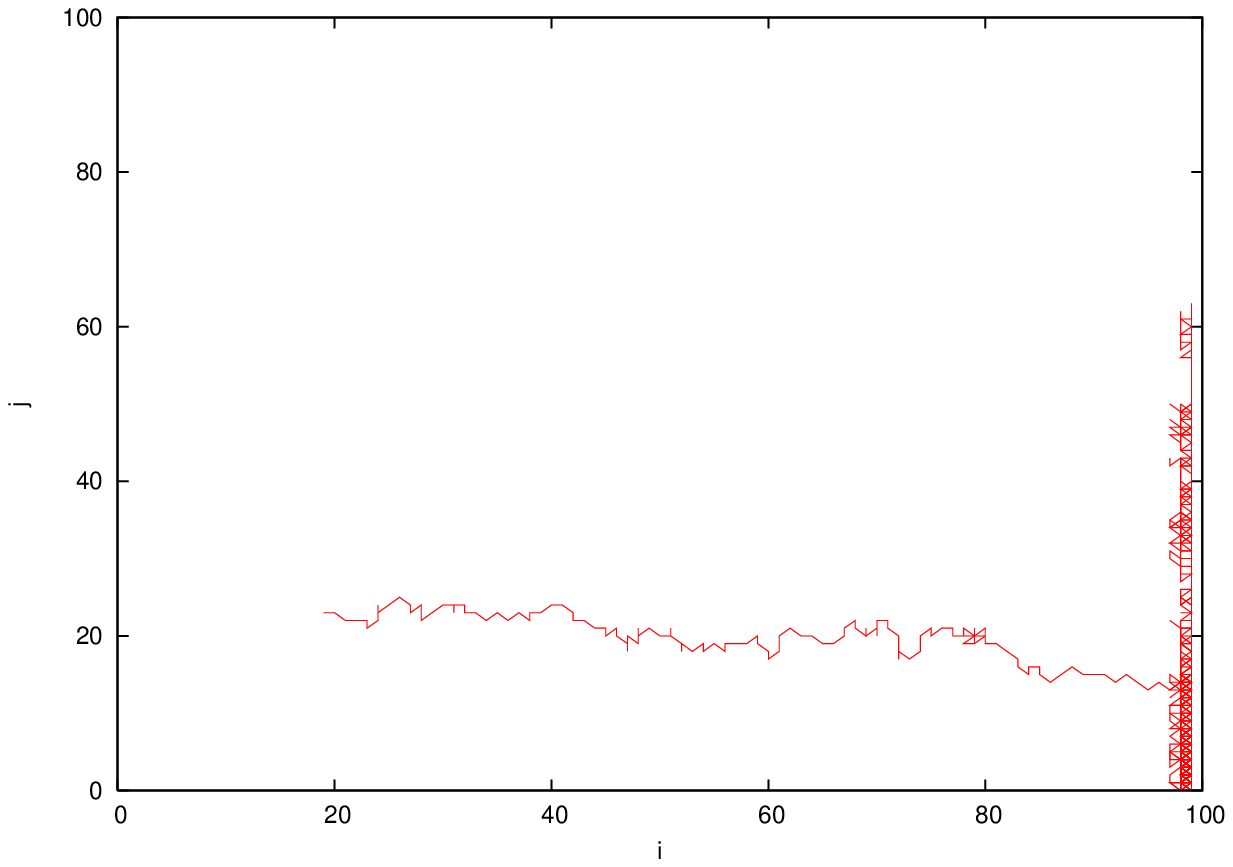}   \\
\includegraphics[width=6cm]{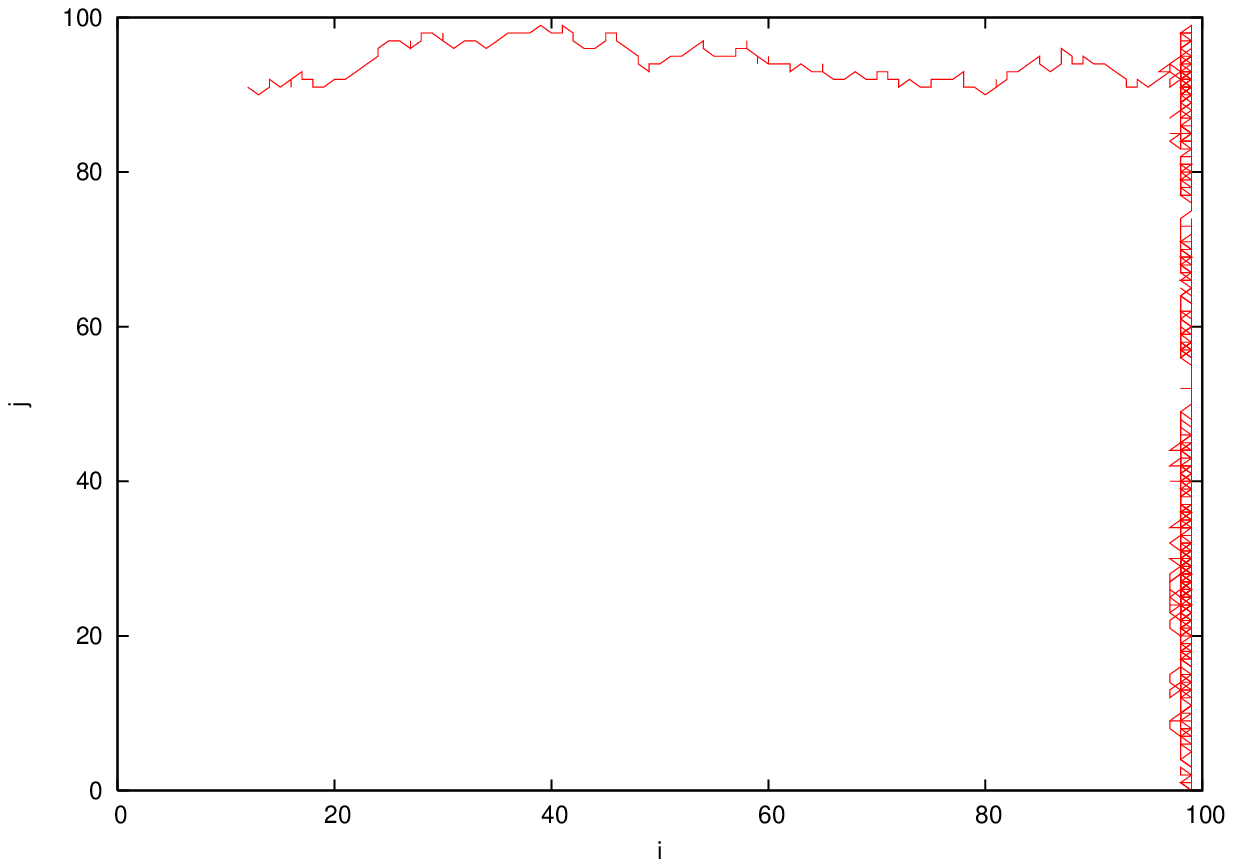}  &  \includegraphics[width=6cm]{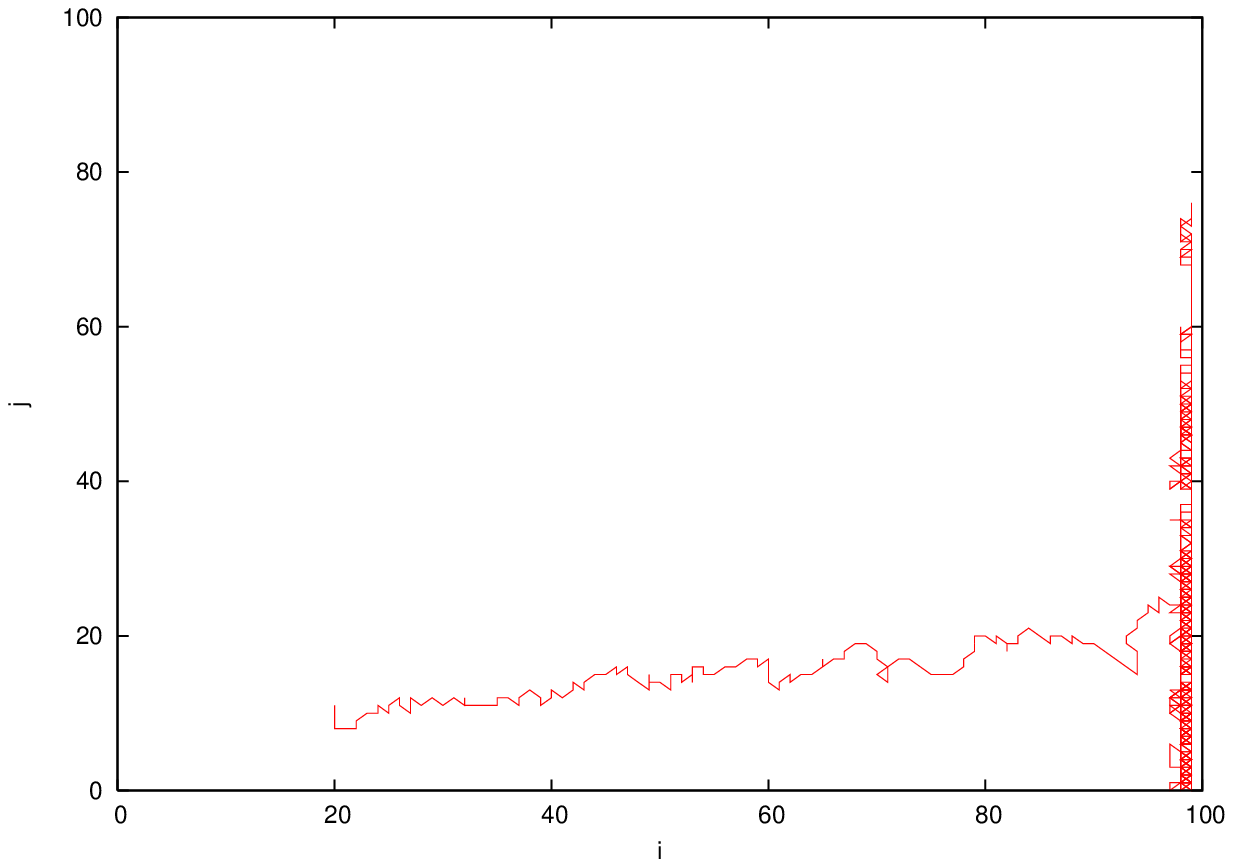}
\end{array}
$\caption{
{\bf Top left}: $\lambda=100$.
{\bf Top right}: $\lambda=100$.
{\bf Bottom left}:  $\lambda=100$.
{\bf Bottom right}:  $\lambda=100$.}
\label{erg}
\end{figure}

The very important question related to Monte Carlo simulations is the
"ergodicity" of the algorithm. It means that in the finite number of steps
(finite time) the system must be freely close to any point in the phase
space. This prevents the system to being trapped in a subset of states.
In the Monte Carlo simulations it causes that we can always find a proper
energy minimum, even for very low temperatures when fluctuations are small.
To check it we performed Markov chains in the low temperature system. In such
a system, when algorithm is not ergodic, a Markov chain cannot always lead to
the proper minimum. In Fig.~\ref{erg} we show Markov chain in the parameters
space for the third model considered in Sec.~\ref{SimpleExample}. We show
that starting from the different points in the parameter space system always
go to the same region where the proper minimum is placed. This is a visual
proof of ergodicity for a kind of function considered. It is possible that it
is not true for more complicated kind of functions.

\end{document}